\documentclass[10pt,a4paper]{article}

\usepackage{amsmath}
\usepackage{hyperref}
\usepackage{graphicx}
\usepackage{epsfig,latexsym,amssymb,color,a4wide,amsfonts,amsbsy}
\usepackage{color}
\usepackage[latin1]{inputenc}

\newcommand{\be}{\begin{equation}}
\newcommand{\ee}{\end{equation}}
\newcommand{\beann}{\begin{eqnarray*}}
\newcommand{\eeann}{\end{eqnarray*}}
\newcommand{\bea}{\begin{eqnarray}}
\newcommand{\eea}{\end{eqnarray}}
\newcommand{\lb}{\label}
\newcommand{\bdm}{\begin{displaymath}}
\newcommand{\edm}{\end{displaymath}}
\newcommand{\D}{{\rm d}}
\newcommand{\E}{{\rm e}}
\newcommand{\I}{{\rm i}}

\begin{document}

\begin{titlepage}

\title{Indefinite oscillators and black-hole evaporation}

\author{%
Claus Kiefer\thanks{\textsf{kiefer@thp.uni-koeln.de}}\\
Institut f\"ur Theoretische Physik, Universit\"at zu K\"oln,\\
Z\"ulpicher Stra\ss e\ 77, 50937 K\"oln, Germany.\\[3mm]
João Marto\thanks{\textsf{jmarto@ubi.pt}}\;\; and
Paulo Vargas Moniz\thanks{\textsf{pmoniz@ubi.pt}}~\thanks{%
URL: \textsf{http://www.dfis.ubi.pt/$\sim$pmoniz}}~\thanks{%
Also at CENTRA, IST, Rua Rovisco Pais, 1049 Lisboa Codex, Portugal.}\\
Departamento de Fisica,
Universidade da Beira Interior, \\ Rua Marqu\^es d'Avila e Bolama,
6200 Covilh\~a, Portugal.}
\date{}
\maketitle
\begin{abstract}
We discuss the dynamics of two harmonic oscillators of which
one has a negative kinetic term. This model mimics the Hamiltonian in
quantum geometrodynamics, which possesses an indefinite kinetic
term. We solve for the time evolution in both the uncoupled and
coupled case. We use this setting as a toy model for studying some
possible aspects of the final stage of black-hole evaporation. We
assume that one
oscillator mimics the black hole, while the other mimics Hawking radiation. In
the uncoupled case, the negative term leads to a squeezing of the
quantum state, while in the coupled case, which includes back
reaction, we get a strong entangled state between the 
mimicked black hole and the
radiation. We discuss the meaning of this state.
We end by analyzing the limits of this model and its
relation to more fundamental approaches.
\end{abstract}

\end{titlepage}


\section{Introduction}

In spite of many attempts, a final theory of quantum gravity
remains elusive \cite{OUP}. Such a theory is, however, needed for
at least two reasons. First, there are conceptual and formal arguments which
point to an encompassing fundamental framework. And second, such a theory
is required in order to tackle concrete physical problems. Among the
most important ones are problems in cosmology and black-hole physics.
In our paper we address the latter field. There, two main issues
are usually studied in this context: the microscopic interpretation
of black-hole entropy and the description of the final evaporation
phase. The first problem is already relevant for large black holes, that is,
for black holes with a mass much bigger than the Planck mass. Therefore, some
progress has been achieved in various approaches such as string theory,
loop quantum gravity, and quantum geometrodynamics
\cite{OUP,Rovelli,Zwiebach}. On the other hand, as the black hole
approaches the Planck mass, 
a full understanding of the theory is necessary for the final
evaporation phase, 
and it is therefore not surprising that no definitive conclusion has
been reached up to now.

In our paper we attempt to shed some light on the question
of how the final phase in the black-hole evolution may be described in
terms of quantum states used in 
canonical quantum gravity. In this approach, the central kinematical
entities are
states that depend -- in the gravitational sector -- on the
three-metric (in quantum geometrodynamics), a non-abelian connection
(in quantum connection dynamics), or a SU(2)-holonomy
(in loop quantum gravity) \cite{OUP,Rovelli}.
The central `dynamical' equations are the Hamiltonian and the
diffeomorphism constraints (plus, in the latter two versions, the
Gauss constraint). The final stage of black-hole evaporation has not
yet been described in this language. There exist several attempts
using the semiclassical Einstein equations, $G_{\mu\nu}=8\pi G
\langle T_{\mu\nu}\rangle$, in order to implement the back reaction
of the Hawking radiation on the evaporating black hole, with no final result.
Since the semiclassical Einstein equations cannot be fundamentally correct
\cite{OUP}, a definite answer can only be obtained from the
exact theory. It is the purpose of this paper to provide a possible
insight into the quantum states of the last evaporation stage.
We are certainly not able to
present a realistic description of the final stage.
But we can address and
tentatively answer the following question: Suppose we have solved for
the quantum 
state -- how can we recognize black hole evaporation from it?
We achieve this goal through some admittedly oversimplified models which,
however, as we hope, would capture some real features of the physical process
under investigation. The model is mainly chosen in order to study the
influence of an indefinite kinetic term, as it occurs in the
Wheeler--DeWitt equation. 
Thus, in this respect our model may exhibit realistic features of the
black-hole evaporation process, while in other respects it may be
totally unrealistic and must be replaced with features from more realistic
(at the moment not exactly soluble) models. We herein envisage the black hole
as being embedded as a quantum object into a semiclassical
universe. We thus have to address the time-dependent Schr\"odinger
equation with the full quantum gravitational black hole Hamiltonian
mimicking the black hole. 

Our paper is organized as follows. Section~2 presents a brief review
of the semiclassical expansion. We present our arguments why we use a
(functional) Schr\"odinger equation with a quantum gravitational
Hamiltonian aiming to describe a quantum black hole.
 Section~3 is the main part of our paper. We
present our model of coupled oscillators mimicking the quantum black
hole and Hawking radiation. We first consider the case without direct
back reaction (only indirectly through the decrease of the black-hole
mass). The negative kinetic term 
(characteristic of a black-hole Hamiltonian) leads
to a squeezed state when the mass which we associate with a black hole
approaches the Planck
mass. We then discuss the case of a direct coupling. This leads to an
entangled state between the mimicked black hole and the Hawking radiation. We
calculate and discuss the corresponding reduced density
matrices. Section~4 contains our conclusions.


\section{Semiclassical limit and beyond}

The central equations of canonical quantum gravity in the
geometrodynamical, connection, or loop approach are of the
constraint form $H\Psi=0$. We shall restrict in our paper attention to
quantum geometrodynamics; from loop quantum gravity one would expect
in addition some features arising from the discreteness of space, which
are not captured here, but
which could play a crucial role in a more realistic scenario
\cite{Rovelli}. The first point to notice is that the constraint
equations in the form $H\Psi=0$
are not yet suitable to describe black-hole
evaporation. This is because these equations, as they stand,
describe a closed system where everything is described quantum gravitationally.
In order words, they would describe the case of a quantum black hole
within a quantum universe. Although this may be the appropriate picture
at the most fundamental level, the situation that one wants to address
is a quantum black hole embedded in a {\em semiclassical} universe for
which an appropriate time parameter is present.

Such a time variable can arise in various ways. One can, for example,
discuss the gravitational collapse of a dust cloud, in which a dust
proper time emerges in a natural way, cf. \cite{KMV} and the
references therein. Solutions of quantum geometrodynamics can be
exploited to derive Hawking radiation and greybody corrections within
this model \cite{Vaz}. Another way of recovering time is a
Born--Oppenheimer type of expansion scheme at the full (formal) level
of quantum geometrodynamics \cite{OUP}. This is the framework which we
shall use here.

Performing, for example, an expansion with respect to the Planck mass,
one can derive from the full constraint equation (the Wheeler--DeWitt
equation) a functional Schr\"odinger equation for `matter' fields
in an external spacetime. The time in this Schr\"odinger equation
is of a semiclassical nature and defined by configurations of the
semiclassical gravitational field. More concretely, this equation reads
\begin{eqnarray}
        \I\hbar\frac{\partial}{\partial t}\,
        |\psi(t)\rangle &=& \hat{H}{}^{\rm m}|\psi(t)\rangle\ ,\nonumber \\
        \hat{H}{}^{\rm m} &\equiv&
        \int \D^3 x \left\{N({\bf x})
        \hat{\mathcal H}{}^{\rm m}_{\perp}({\bf x})+
        N^a({\bf x})\hat{\mathcal H}{}^{\rm m}_a({\bf x})\right\}\ .
       \label{semi}
        \end{eqnarray}
Usually, $\hat{H}{}^{\rm m}$ is the Hamiltonian for the `matter'
(i.e. non-gravitational) fields in the Schr\"odinger
picture, parametrically depending on (generally non-static) metric
coefficients of the curved space--time background. The `bra' and the `ket'
in the quantum state refer to the Hilbert space of the `matter'
fields. ($N$ and $N^a$ are lapse and shift function, respectively,
and $\hat{\mathcal H}{}^{\rm m}_{\perp}({\bf x})$ and
$\hat{\mathcal H}{}^{\rm m}_a({\bf x})$ denote the quantum `matter'
part of the Hamiltonian
constraint and diffeomorphism constraint operator, respectively.)
It is important to emphasize that only the semiclassical degrees of freedom
of the gravitational field enter the definition of $t$.

In realistic situations, the full set of
these semiclassical variables may, however, include only part of the
gravitational field and may also include part of the matter. On the
other hand, 
the gravitons, which are small excitations of the metric, behave fully
quantum and must be included into $\hat{H}{}^{\rm m}$.
But this is also what happens in our case here: {\em The degrees of freedom
corresponding to the quantum black hole
must be part of
$\hat{H}{}^{\rm m}$}, while the time parameter $t$ is defined by the
macroscopic part of gravity and matter, that is, by the semiclassical
Universe into which the quantum black hole is embedded.

To describe black-hole evaporation, therefore, one should stick to
this semiclassical description of the embedding universe -- providing
the time parameter $t$ --,
while employing a full quantum description for the black hole.
This corresponds to the realistic situation of an observer
residing outside the quantum black hole and having the usual semiclassical
time at his disposal. In such a situation one has therefore
to apply Equation \eqref{semi} with $t$ referring to the semiclassical
time of the outside universe, and $\hat{H}{}^{\rm m}$ being the full Hamilton
operator of the {\em quantum} black hole {\em and} the fields
interacting with it. It is for this reason that we consider an
indefinite kinetic term in $\hat{H}{}^{\rm m}$; it is inherited from
the Wheeler--DeWitt equation and applicable here because we apply
quantum gravity to the black hole. 
One thus has to employ a mixture of Schr\"odinger and Wheeler--DeWitt
equation.\footnote{The quantum {\em formation} of a black hole from
  spherical domain-wall collapse was studied recently in a related
  framework in \cite{VSK}.}

The kinetic term of the gravitational part of the quantum Hamilton
operator is suppressed by the Planck mass, $m_{\rm P}$. As long as the
black-hole mass
is large, this kinetic term should thus be irrelevant. 
One has in this limit the Hawking radiation as
the only contribution to \eqref{semi}. The corresponding quantum state was
explicitly calculated and discussed in \cite{DK}.
After the black hole approaches the Planck mass in the final
evaporation phase, the full quantum Hamiltonian
of the black hole becomes relevant. In the full field theoretic framework
of \cite{DK}, this was not considered. We shall instead investigate this
question here in a quantum mechanical model that can capture {\em some}
of the relevant features. In particular, as emphasized above, we have
to deal with the 
important property of the gravitational Hamiltonian possessing an
indefinite kinetic term. This is important for the understanding of
time in quantum gravity \cite{OUP,Zeh}, but will also be crucial
for the qualitative features of black-hole evaporation.


\section{A simple model of black-hole evaporation}

The full equation \eqref{semi} is a complicated functional
differential equation for a wave functional depending on the
three-metric and matter fields (including Hawking radiation). In order
to make our analysis tractable, we shall instead consider the
following model, which is purely quantum mechanical:
\begin{eqnarray}
\lb{model}
\I\hbar\frac{\partial }{\partial t}\Psi (x,y,z,t) &=&\left(\frac{\hbar^2}
{2m_{\rm P}}%
\frac{\partial ^{2}}{\partial x^{2}}-\frac{\hbar^2}{2m_{y}}\frac{\partial ^{2}}{%
\partial y^{2}}-\frac{\hbar^2}{2m_{z}}\frac{\partial ^{2}}{\partial
z^{2}}\right. +
\label{model} \\
&&\left. \frac{m_{\rm P}\omega _{x}^{2}}{2}x^{2}+\frac{m_{y}\omega _{y}^{2}}{2}%
y^{2}+\frac{m_{z}\omega _{z}^{2}}{2}z^{2}\right) \Psi (x,y,z,t)\ .  \nonumber
\end{eqnarray}
How do we interpret this ansatz? Let us first recall that the time $t$
in this equation is the semiclassical time of \eqref{semi} coming from
the semiclassical degrees of freedom (such as the scale factor and
macroscopic matter) of the Universe.
Equation \eqref{semi}
contains second functional derivatives with respect to the
three-metric describing the black hole. We shall mimic this
three-metric by the single variable $x$; we expect $x$ to be the
mass, $M$, of a Schwarzschild black hole (more precisely, its
Schwarzschild radius $2GM/c^2$). The `wrong' sign of its kinetic term
reflects this correspondence to the three-metric. 
At a stage where the black hole is quite large, we expect that
the kinetic term referring to $x$ is negligible. However, for the
final phase which we shall discuss in this paper, $M$ will be
small. Therefore, such a term will be of relevance, which is why it
has been included into \eqref{model}.

The variable $y$ is supposed to be the analogue to Hawking radiation,
with $m_y$ corresponding to its energy, and
$z$ stands for the remaining degrees of freedom (with $m_z$ as a
formal mass parameter).
 We shall consider
in the following only the $x$- and $y$-variables, that is, we shall
restrict attention to the black hole and its interaction with
Hawking radiation. We assume that at least for large black holes the
evolution is stable in the sense that the potential does not become
too negative. For simplicity we have chosen harmonic oscillator
potentials; this is realistic for the description of the Hawking
radiation \cite{DK}, but definitely oversimplified for the black hole.
In \eqref{model} we have not yet taken into account the direct back
reaction of Hawking radiation ($y$-part) onto the black hole
($x$-part), that is, no $x$-$y$-coupling is included in
\eqref{model}. This back reaction will be implemented in Section~3.3
below.

We admit that the whole ansatz is not fully realistic, but
we hope that our results will shed at least some light on the {\em
interpretation} of black-hole evaporation in quantum gravity: the
setting associated with (2) is not that of a standard quantum mechanical
description of harmonic oscillators. On the one hand, there is an
unusual kinetic term. As  we shall see, the main consequences in our
{\em analogue} to black-hole evaporation come from the negative kinetic
term for $x$ and are independent of the details of the potential. On
the other hand, the Planck mass appears as a parameter (suggested
from the  full Wheeler-DeWitt equation, where it appears in front of the
derivatives with respect to the three-metric), participating in 
(2) as a natural `suppressor'; this is different from the dynamics of
the Hawking radiation \cite{DK}.

Let us be more precise, taking only the $x$- and
$y$-degrees of freedom into account, one arrives at a Schr\"odinger
equation that can be solved by a separation ansatz, \be \Psi(x,y,t)=
\psi_x(x,t)\psi_y(y,t)\ , \ee with (absorbing the separation
constant into a state redefinition) \bea \I\hbar\dot{\psi_x}(x,t)
&=& \left(\frac{\hbar^2}{2m_{\rm P}} \frac{\partial ^{2}}{\partial
x^{2}}+\frac{m_{\rm P}\omega _{x}^{2}}{2}x^{2}
\right)\psi_x(x,t)\ , \lb{psix} \\
\I\hbar\dot{\psi_y}(y,t) &=& \left(-\frac{\hbar^2}{2m_y}
\frac{\partial ^{2}}{\partial y^{2}}+\frac{m_y\omega _{y}^{2}}{2}y^{2}
\right)\psi_y(y,t)\ \lb{psiy} .
\eea
We see from the first of these equations that $\psi_x^*$ obeys a
Schr\"odinger equation with standard kinetic term, but with the
sign of the potential being {\em reversed}
(`upside-down oscillator'). This will become important
in the following discussion. Wave functions describing the Schwarzschild
and the Reissner--Nordstr\"om black hole have been discussed
in quantum geometrodynamics at various places, see for example
\cite{OUP,kuchar,LWH,KL,BK}. This part of the total state is mimicked
by the $x$-system. Our purpose here is to construct Gaussian wave packets
which are solutions of \eqref{psix} and \eqref{psiy},
 and which describe the transition
from the semiclassical regime to the final evaporation phase.
For various relevant aspects of such wave packets in quantum mechanics, cf.
\cite{Rob1,Rob2,andrews}.

If $\psi_0(x,0)$ denotes an initial state, the evolution
for the $x$-part is found from
 \begin{equation}
\int dx^{\prime }\ G(x,x^{\prime };t,0)\psi _{0}(x^{\prime },0)=\psi
(x,t)\ ,  \label{eq.7}
\end{equation}
where $G(x,x^{\prime };t,0)$ denotes the Green function. In the case of the inverted oscillator
it reads (\cite{GS}, Sec. 6.2.1.8):
\begin{equation}
G(x,x^{\prime };t,0)=\sqrt{\frac{m_{\rm P}\omega _{x}}{2\pi \I\hbar
\sinh (\omega _{x}t)%
}}\exp \biggl[\I m_{\rm P}\omega _{x}\frac{(x^{2}+x^{\prime }{}^{2})\cosh (\omega
_{x}t)-2xx^{\prime }}{2\hbar\sinh (\omega _{x}t)}\biggr]\ .  \label{eq.n12}
\end{equation}
We shall now investigate the solutions of the Schr\"odinger equation
for various initial states.

\subsection{Squeezed Ground State}

Before we address the more interesting case of an initial coherent
state in Sec.~3.2,
we take in a first example as an initial state the ground state of the
harmonic oscillator,
\be
\psi _{x0}^{g}(x^{\prime },0)=\left(\frac{m_{\rm
      P}\omega_x}{\pi\hbar}\right)^{1/4}
\exp\left(-\frac{m_{\rm P}\omega_x}{2\hbar}x^2\right)\ .
\ee
For an ordinary oscillator, the system would stay in the ground
state. Here, however, we find from
\eqref{eq.7}, together with the complex conjugation,
cf. \eqref{psix}, the state
\be
\lb{ground}
\psi_x^{g}(x,t)=\left(\frac{m_{\rm P}\omega_x}{\pi\hbar(1-\I\sinh 2\omega_x t)}
\right)^{1/4}\exp\left(-\frac{m_{\rm P}\omega_x}{2\hbar\cosh 2\omega_xt}
(1+\I\sinh2\omega_xt)x^2\right)\ .
\ee
This is, in fact, a squeezed ground state.
 Comparing with the general form
of Gaussian squeezed states, see for example \cite{Nieto},
one recognizes that the squeezing angle is $\phi=\pi/4$, and the squeezing
parameter is $r=\omega_xt$. The squeezing thus proceeds along the
diagonal in phase space and increases linearly with time.
Recalling the analogy to the black-hole case, this squeezing is
expected to happen
in the final evaporation phase when the kinetic term for the
$x$-degree of freedom becomes significant.
The squeezing vanishes in
the formal limit $m_{\rm P}\to\infty$. The evolution of the $x$-part
of the wave packet is depicted in Figure~1.

\begin{figure}[h]
\setlength{\belowcaptionskip}{15pt}
\renewcommand{\baselinestretch}{1.2}
  \centering \includegraphics[width=5.5in]{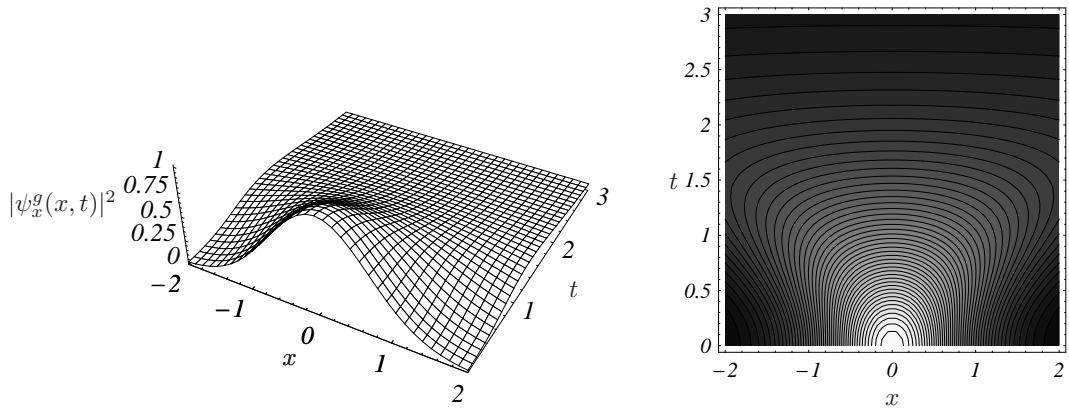}
 \caption{{\protect \footnotesize Evolution of a Gaussian state under
 the inverted oscillator propagator. We depict $|\psi_x^{g}(x,t)|^2$ with
 $m_{\rm P}=\hbar=\omega_x=1$ for simplicity. In the contour plot the
 brighter areas correspond to higher values for
 $\vert\psi_x^{g}(x,t)\vert^2$.}}
\end{figure}

Assuming also for the $y$-part the ground state as initial condition,
it is obvious that it will remain in this state if the standard
harmonic oscillator propagator is used. The full state is thus
the product of \eqref{ground} for the $x$-part with the standard
ground state for the $y$-part.

The evaporation of a black hole would thus be mimicked by the
squeezing of a Gaussian wave packet. It is known that Hawking
radiation is in a squeezed state for semiclassical black holes
\cite{GS,CK01}. Here, instead, we have a squeezing of the black-hole
quantum state itself.

This case of a squeezed ground state is unrealistic because the
support of the wave function in the
$x$-variable extends significantly into the negative regime.
Therefore we shall turn to a
coherent state as an appropriate initial condition.

\subsection{Initial Coherent State}

We shall now consider an initial coherent state
for both the $x$- and the $y$-part. This choice captures more the
idea of an initial semiclassical black hole in which the dynamical
variable $x$ may be assumed to be concentrated around an initial
mass $M$. Addressing first the $x$-part, we have
\be \lb{initial2}
\psi_{x0}^{\alpha}(x,0)=\left(\frac{m_{\rm
P}\omega_x}{\pi\hbar}\right)^{1/4} \exp\left(-\frac{m_{\rm
      P}\omega_x}{2\hbar}x^2+\alpha\sqrt{\frac{2m_{\rm P}\omega_x}
{\hbar}}x-\frac{\vert\alpha\vert^2}{2}-\frac{\alpha^2}{2}\right)\ ,
\ee
where
\be
\alpha=\sqrt{\frac{m_{\rm P}\omega_x}{2\hbar}}x_0+\I\frac{p_0}
{\sqrt{2m_{\rm P}\omega_x}}\ ,
\ee
with $x_0$ and $p_0$ denoting the expectation values of $x$ and $p_x$,
respectively. The state \eqref{initial2} thus contains two free
parameters $x_0$ and $p_0$, which together form the complex variable
$\alpha$.
Application of \eqref{eq.n12}
and complex conjugation then gives
\bea
\lb{coherent}
& & \psi_x^{\alpha}(x,t)=\left(\frac{m_{\rm P}\omega_x}{\pi\hbar(1-\I\sinh 2\omega_x t)}
\right)^{1/4}\exp\left(-\frac{m_{\rm P}\omega_x}{2\hbar\cosh 2\omega_xt}
(1+\I\sinh2\omega_xt)x^2\right)\times \nonumber\\
& & \quad \exp\left(\I\alpha^*\sqrt{\frac{2m_{\rm
P}\omega_x}{\hbar}} \frac{\sinh\omega_xt-\I\cosh\omega_xt}{\cosh
2\omega_xt}x -\frac{\vert\alpha\vert^2}{2}-\frac{\alpha^{*2}}{\cosh
2\omega_xt} (\frac12+\I\cosh\omega_xt)\right) . \eea Comparing
\eqref{coherent}
with \eqref{ground}, we immediately recognize that this state
experiences the same degree of squeezing. The state \eqref{ground}
is, of course, a special case of \eqref{coherent} from which it
follows for $\alpha=0$. The absolute square of \eqref{coherent} is
depicted in Figure~2. 

Our model allows both signs for $p_0$. In order to capture the
idea of an initially decreasing mass we choose as initial
condition a negative value for $p_0$. This is shown in Figure~2.
One can recognize that
the localization of the packet centre moves toward smaller values of $x$
before it spreads. The spreading in this model should reflect the
quantum gravitational behaviour in the realistic situation.
In our simplified model it will be possible that also
negative values of $x$ can be reached, which is not possible if $x$
represents the Schwarzschild radius; in a more realistic model, a
potential wall at $x=0$ should be introduced. 

\begin{figure}[h]
\setlength{\belowcaptionskip}{15pt}
\renewcommand{\baselinestretch}{1.2}
  \centering \includegraphics[width=5.5in]{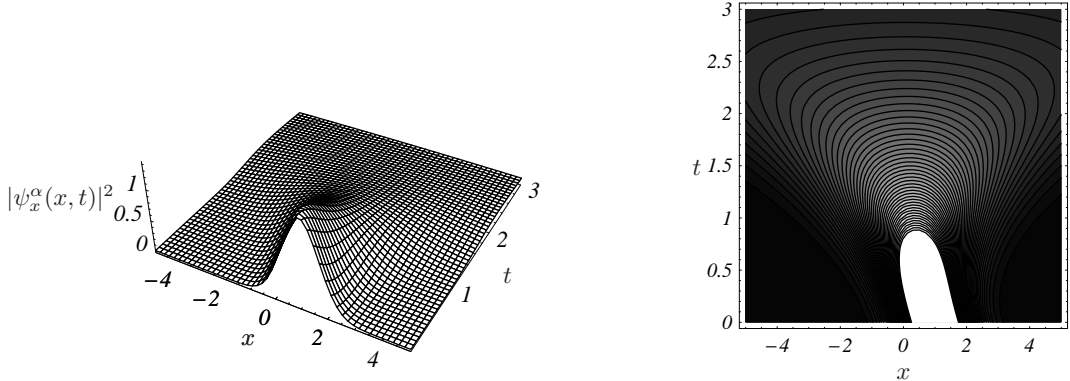}
 \caption{{\protect \footnotesize Evolution of
     $\vert\psi_x^{\alpha}(x,t)\vert^2$ under
 the inverted oscillator propagator, where
 $m_{\rm P}=\hbar=\omega_x=x_{0}=1$ for simplicity and with $p_{0}=-1$.
 In the contour plot the brighter areas correspond to higher values
 for $\vert\psi_x^{\alpha}(x,t)\vert^2$.}}
\end{figure}

The absolute square of \eqref{coherent} can also be written in the form
\bea
\lb{coh1}
\vert\psi_x^{\alpha}(x,t)\vert^2&=& \left(\frac{m_{\rm
      P}\omega_x}{\pi\hbar\cosh 2\omega_xt}
\right)^{1/2}f(x_0,p_0,t)\times\nonumber\\
& & \ \exp\left(-\frac{m_{\rm P}\omega_x}{\hbar\cosh 2\omega_xt}
(x-[x_0\cosh \omega_xt+\frac{p_0\sinh\omega_xt}{m_{\rm
    P}\omega_x}])^2\right)\ ,
\eea
where the explicit form of $f(x_0,p_0,t)$ is of less interest. It is easily
seen from this result that the packet is peaked around the classical
solution, but highly squeezed. We note that this fact makes it very
sensitive to decoherence, see Section~4.

Taking for the $y$-system (the ordinary oscillator) an initial coherent
state as in \eqref{initial2}, one obtains the standard result
for the time-dependent coherent state, as found in textbooks on
quantum mechanics,
\bea
\lb{coherentnormal}
\psi_y^{\alpha}(y,t)&=&\left(\frac{m_y\omega_y}{\pi\hbar}\right)^{1/4}
\exp\left(-\frac{\I\omega_yt}{2}\right)
\exp\left(-\frac12\left[\sqrt{\frac{m_y\omega_y}
{\hbar}}y-\sqrt{2}\alpha\E^{-\I\omega_yt}\right]^2\right)\nonumber\\
& & \times \exp\left(-\frac12\left[\vert\alpha\vert^2-\alpha^2\E^{-2\I\omega_y
t}\right]\right)\ .
\eea
The time evolution of the absolute square is shown in Figure~3.
(More realistically, one should take into account a mild squeezing for
this state, since it describes Hawking radiation \cite{CK01}.)
The total state is then again a product of \eqref{coherent}
and \eqref{coherentnormal}.

\begin{figure}
\setlength{\belowcaptionskip}{15pt}
\renewcommand{\baselinestretch}{1.2}
  \centering \includegraphics[width=5.5in]{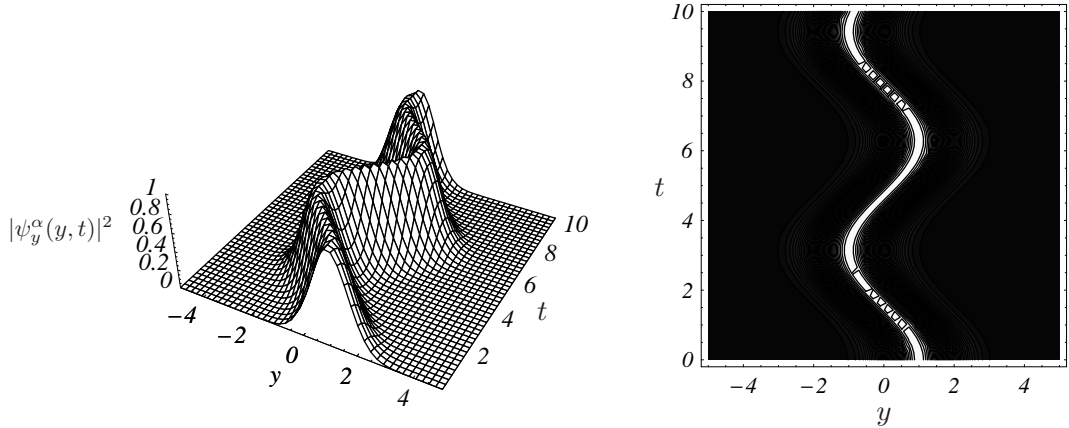}
 \caption{{\protect \footnotesize Evolution of
     $\vert\psi_y^{\alpha}(y,t)\vert^2$ under
 the ordinary oscillator propagator, with
 $m_{y}=\hbar=\omega_y=y_{0}=p_{0y}=1$ for simplicity. }}
\end{figure}

\subsection{Back reaction between a black hole and Hawking radiation}

In the previous section, back reaction was only implicitly implemented
in the sense that the kinetic term was assumed to become relevant for small
black holes.
In order to be more realistic, a direct coupling between the black hole
and its Hawking radiation, that is, a coupling between the $x$-and $y$-part,
has to be included. This will mimic the process of back reaction more
clearly.
The simplest way is to include into \eqref{model} a linear coupling
of the form $\mu xy$, where the parameter $\mu$ should depend on the
black-hole parameters. Then equation \eqref{model}, when considering only
the $x$-and $y$-part, will transform into
\begin{eqnarray}
\lb{model1}
\I\hbar\frac{\partial }{\partial t}\Psi (x,y,t) &=&\left(\frac{\hbar^2}
{2m_{\rm P}}%
\frac{\partial ^{2}}{\partial x^{2}}-\frac{\hbar^2}{2m_{y}}\frac{\partial ^{2}}{%
\partial y^{2}}\right. +
\label{eq.6} \\
&&\left. \frac{m_{\rm P}\omega _{x}^{2}}{2}x^{2}+\frac{m_{y}\omega _{y}^{2}}{2}%
y^{2}+ \mu xy\right) \Psi (x,y,t)\ .  \nonumber
\end{eqnarray}
What would be a suitable initial state? As for the $x$-part, one
could still start with an initial coherent state. As for the
$y$-part, mimicking Hawking radiation, we would suggest to take the
following state as initial state,
\be
\lb{Hawking}
\psi_{y0}^{H}(y,t_0)\propto \exp\left(-\frac{m_y\omega_y}{2\hbar}
{\rm coth}\left[\frac{2\pi\omega_yGM}{c^3}+\I\omega_yt_0\right]y^2\right)\ ,
\ee
where $M$ is the original mass of the (Schwarzschild) black hole, which
corresponds to the initial value $x_0$ of the $x$-part of the quantum
state.
What is the justification for this choice? As was shown in \cite{DK},
\eqref{Hawking} corresponds to the state describing Hawking
radiation. In \cite{DK}
we have dealt with a dilaton model; for the Schwarzschild case one would
expect the form \eqref{Hawking} of the state, cf. \cite{CK01}.

It is appropriate to rewrite Equation \eqref{model1} in the following way:
\be
\lb{simpl1}
\I\hbar\frac{\partial }{\partial t}\Psi (Q_{1},Q_{2},t)
=\biggl[\frac{1}{2}\left(P_{2}^2-P_{1}^2\right)+
\frac{1}{2}\left(\Omega_{1}^2 Q_{1}^2 +\Omega_{2}^2 Q_{2}^2
\right)+\Gamma Q_{1}Q_{2}\biggr]\Psi (Q_{1},Q_{2},t) ,
\ee
in which the following redefinitions are used:
\begin{align}
    x&=
    \frac{Q_{1}\cos{\theta}+Q_{2}\sin{\theta}}{\sqrt{m_{\rm P}}\cos{2\theta}} &&
    P_x=
    \sqrt{\frac{m_{\rm P}}{\cos{2\theta}}}
    \biggl(P_{1}\cos{\theta}+P_{2}\sin{\theta}\biggr) \nonumber \\
    y&=\frac{Q_{1}\sin{\theta}+Q_{2}\cos{\theta}}{\sqrt{m_y}\cos{2\theta}} &&
    P_y=
    \sqrt{\frac{m_{y}}{\cos{2\theta}}}
    \biggl(P_{1}\sin{\theta}+P_{2}\cos{\theta}\biggr) \nonumber \\
    \Omega_{1}^2\cos^2{2\theta}&=\omega_{x}^2\cos^2{\theta}+
        \omega_{y}^2\sin^2{\theta}+
        \frac{\mu\sin{2\theta}}{\sqrt{m_{\rm P}m_{y}}}  
    &&\widetilde{\omega}_{y}=\frac{\omega_{y}}{\cos^2{2\theta}}
    {\rm coth}\left[\frac{2\pi\omega_yGM}{c^3}+\I\omega_yt_0\right]\nonumber \\
    \Omega_{2}^2\cos^2{2\theta}&=
    \omega_{x}^2\sin^2{\theta}+\omega_{y}^2\cos^2{\theta}+
    \frac{\mu\sin{2\theta}}{\sqrt{m_{\rm P}m_{y}}} 
    &&\widetilde{\omega}_{x}= \frac{\omega_{x}}{\cos^2{2\theta}} .
  \lb{transf1}
\end{align}
Here, without any loss of generality,
\be
\Gamma = \frac{1}{\cos^2{2\theta}}\biggl(\frac{1}{2}(\omega_{x}^2+
\omega_{y}^2)\sin{2\theta}+\frac{\mu}{\sqrt{m_{\rm
      P}m_{y}}}\biggr)
\ee
must be zero in order that \eqref{simpl1} allows a separation of
variables \cite{Ben}. In this case we find
\be
\lb{mu1}
\mu=-\frac{1}{2}
\sqrt{m_{\rm P}m_{y}}(\omega_{x}^2 +\omega_{y}^2)\sin{2\theta} 
\ee
with $\theta\in ]-\frac{\pi}{4},\frac{\pi}{4}[$ in order to have the set 
of transformations \eqref{transf1} physically consistent. It
is important to point out that $\mu$ can assume any 
value in the interval $\theta\in ]-\frac{\pi}{4},\frac{\pi}{4}[$.

 From \eqref{transf1} and \eqref{mu1} we obtain the following results,
\begin{align}
    \Omega_{1}^2&=\frac{1}{\cos^2{2\theta}}\biggl[\omega_{x}^2\biggl(\cos^2{\theta}-\frac{1}{2}\sin^2{2\theta}\biggr)+
                  \omega_{y}^2\biggl(\sin^2{\theta}-\frac{1}{2}\sin^2{2\theta}\biggr)\biggr]\
                  ,\nonumber \\
    \Omega_{2}^2&=\frac{1}{\cos^2{2\theta}}\biggl[\omega_{x}^2\biggl(\sin^2{\theta}-\frac{1}{2}\sin^2{2\theta}\biggr)+
                   \omega_{y}^2\biggl(\cos^2{\theta}-\frac{1}{2}\sin^2{2\theta}\biggr)\biggr].\lb{omegas}
\end{align}

We are now in a position to apply the same procedures as in Section 3.
We must use, of course, the form of the initial states
\eqref{initial2} and \eqref{Hawking} under the redefinitions
\eqref{transf1}. In order to obtain the time evolution, the next step
consists in calculating the following integral,

\begin{align}
\int\int dQ_{1}' dQ_{2}' \quad & G(Q_1,Q_{1}';t)\cdot G(Q_2,Q_{2}';t) \nonumber \\
& \exp\left(-\frac{\widetilde{\omega}_{y}}{2\hbar}
(Q_{1}'^2\sin^2{\theta}+Q_{2}'^2\cos^2{\theta}+Q_{1}'Q_{2}'\sin{2\theta})\right)
\nonumber \\ 
& \left(\frac{m_{\rm P}\omega_x}{\pi\hbar}\right)^{1/4}
\exp\biggl(-\frac{\widetilde{\omega}_x}{2\hbar}(Q_{1}'^2\cos^2{\theta}+Q_{2}'^2\sin^2{\theta} 
+Q_{1}'Q_{2}'\sin{2\theta}) \nonumber\\
&+\alpha^{*}\sqrt{\frac{2\widetilde{\omega}_x}
{\hbar}}(Q_{1}'\cos{\theta}+Q_{2}'\sin{\theta})-
\frac{\vert\alpha\vert^{2}}{2}-\frac{\alpha^{*2}}{2}\biggr), \lb{ev1}
\end{align}
where $G(Q_{1},Q_{1}^{\prime };t)$ is the inverted oscillator propagator and
$G(Q_{2},Q_{2}^{\prime };t)$ the standard one.
After some calculation we obtain
\begin{align}
\psi(Q_{1}, Q_{2}, t)= & \left(\frac{m_{\rm P}\omega_x}{\pi\hbar}\right)^{1/4}
\left(-\frac{\Omega_{1}\Omega_{2}}{{\mathcal{F}}_{1}{\mathcal{F}}_{3}}\right)^{1/2}
\exp{\left[-\left(\frac{Q_{1}^2}{2\hbar}\frac{{\mathcal{F}}_{2}}{{\mathcal{F}}_{1}} 
+\frac{Q_{2}^2}{2\hbar}\frac{{\mathcal{F}}_{4}}{{\mathcal{F}}_{3}}\right)\right]}
\nonumber \\ 
&
\exp{\left[-i\alpha^{*}\sqrt{\frac{2\widetilde{\omega}_x}{\hbar}}
\left(\frac{\Omega_{1}Q_{1}\cos{\theta}}{{\mathcal{F}}_{1}}
+\frac{\Omega_{2}Q_{2}\sin{\theta}}{{\mathcal{F}}_{3}}\right)\right]} \nonumber \\
&
\exp{\left[\frac{\Omega_{2}Q_{2}\sin{2\theta}}{2{\mathcal{F}}_{1}{\mathcal{F}}_{3}}
\left(\frac{\Omega_{1}Q_{1}}{\hbar}-i\alpha^{*}
\sqrt{\frac{2\widetilde{\omega}_x}{\hbar}}\sinh{\Omega_{1}t}\cos{\theta}
\right)\left( \widetilde{\omega}_{y}+\widetilde{\omega}_x \right)\right]} \nonumber \\
&
\exp{\left[-\frac{\hbar \sin^2{2\theta}
\sin{\Omega_{2}t}}{8{\mathcal{F}}_{1}^2 {\mathcal{F}}_{3}}
\left(i\frac{\Omega_{1}Q_{1}}{\hbar}+\alpha^{*}
\sqrt{\frac{2\widetilde{\omega}_x}{\hbar}}\sinh{\Omega_{1}t}\cos{\theta}
\right)^2 \left( \widetilde{\omega}_{y}+\widetilde{\omega}_x \right)^2\right]} \nonumber \\
&
\exp{\left[ \alpha^{*2} \widetilde{\omega}_x\left(\frac{\cos^2{\theta}
\sinh{\Omega_{1}t}}{{\mathcal{F}}_{1}}+
\frac{\sin^2{\theta} \sin{\Omega_{2}t}}{{\mathcal{F}}_{3}}
\right)-\frac{\alpha^{*2}}{2}-\frac{|\alpha|^2}{2}\right]} ,\lb{ev2}
\end{align}
where we have defined the following functions of time,
\begin{align}
{\mathcal{F}}_{1}=&-i\Omega_{1}\cosh{\Omega_{1}t}+\widetilde{\omega}_x\cos^2{\theta}
\sinh{\Omega_{1}t}+\widetilde{\omega}_{y}\sin^2{\theta}\sinh{\Omega_{1}t},
\nonumber \\
{\mathcal{F}}_{2}=&-\Omega_{1}^2\sinh{\Omega_{1}t}-i\Omega_{1}\widetilde{\omega}_x\cos^2{\theta}
\cosh{\Omega_{1}t}-i\Omega_{1}\widetilde{\omega}_{y}\sin^2{\theta}\cosh{\Omega_{1}t} ,
\nonumber \\
{\mathcal{F}}_{3}=&-i\Omega_{2}\cos{\Omega_{2}t}+\widetilde{\omega}_x\sin^2{\theta}
\sin{\Omega_{2}t}+\widetilde{\omega}_{y}\cos^2{\theta}\sin{\Omega_{2}t},
\nonumber\\
&-(\widetilde{\omega}_{y}+\widetilde{\omega}_x)^2\sin^2{2\theta}
\frac{\sinh{\Omega_{1}t}\sin{\Omega_{2}t}}{4{\mathcal{F}}_{1}}
\nonumber \\
{\mathcal{F}}_{4}=&\Omega_{2}^2\sin{\Omega_{2}t}-i\Omega_{2}\widetilde{\omega}_x
\sin^2{\theta}\cos{\Omega_{2}t}-i\Omega_{2}\widetilde{\omega}_{y}\cos^2{\theta}\cos{\Omega_{2}t}
\nonumber\\
& +i\Omega_{2}(\widetilde{\omega}_{y}+\widetilde{\omega}_x)^2\sin^2{2\theta}
\frac{\sinh{\Omega_{1}t}\cos{\Omega_{2}t}}{4{\mathcal{F}}_{1}}.\lb{ft1}
\end{align}
Evaluating the functions ${\mathcal{F}}_{n}$ at $t=0$ enables us to recover the
initial state $\psi(Q_{1}, Q_{2}, t=0)$. Now we must perform the
inverse coordinate transformation,
\be
\begin{cases}
    Q_{1}=\sqrt{m_{\rm P}}x\cos{\theta}-
          \sqrt{m_{y}}y\sin{\theta}\\
    Q_{2}=\sqrt{m_{y}}y\cos{\theta}-
          \sqrt{m_{\rm P}}x\sin{\theta}
\end{cases}
\ee
in order to give the explicit form of $\psi(x, y, t)$. Writing
\be
\lb{sol1}
    \psi(x, y, t)=F(t)\exp{\biggl(A(t)x^2+B(t)x+C(t)y^2+D(x,t)y\biggr)} \ ,
\ee
we find
\begin{align}
F(t)=& \left(\frac{m_{\rm P}\omega_x}{\pi\hbar}\right)^{1/4}
\left(-\frac{\Omega_{1}\Omega_{2}}{{\mathcal{F}}_{1}{\mathcal{F}}_{3}}\right)^{1/2}
\times
\nonumber \\
&
\exp{\left[ \alpha^{*2} \widetilde{\omega}_x
\left(\frac{\cos^2{\theta} \sinh{\Omega_{1}t}}{{\mathcal{F}}_{1}}+
\frac{\sin^2{\theta} \sin{\Omega_{2}t}}{{\mathcal{F}}_{3}} \right)-
\frac{\alpha^{*2}}{2}-\frac{|\alpha|^2}{2}\right]}\times
\nonumber \\
&
\exp{\left[\frac{\alpha^{*2} \widetilde{\omega}_x
(\widetilde{\omega}_{y}+\widetilde{\omega}_x)^2 
\cos^2{\theta}\sin^2{2\theta}\sinh^2{\Omega_{1}t}\sin{\Omega_{2}t}}
{4{\mathcal{F}}_{1}^2{\mathcal{F}}_{3}}\right]},
\nonumber \\
A(t)=&-\frac{m_{\rm P}}{2\hbar}
\biggl[ \frac{{\mathcal{F}}_{2}}{{\mathcal{F}}_{1}}\cos^2{\theta}+ \frac{{\mathcal{F}}_{4}}{{\mathcal{F}}_{3}}\sin^2{\theta}+
\frac{\Omega_{1}\Omega_{2}}{2{\mathcal{F}}_{1}{\mathcal{F}}_{3}}
(\widetilde{\omega}_{y}+\widetilde{\omega}_x)\sin^2{2\theta}
\nonumber \\
&
+\frac{\Omega_{1}^2(\widetilde{\omega}_{y}+
\widetilde{\omega}_x)^2\sin^2{2\theta}\cos^2{\theta}
\sin{\Omega_{2}t}}{4{\mathcal{F}}_{1}^2{\mathcal{F}}_{3}} \biggr],
\nonumber\\
B(t)=&-i\alpha^{*}
\sqrt{\frac{2m_{\rm P}\widetilde{\omega}_x}{\hbar}}\biggl[\frac{\Omega_{1}
\cos^2{\theta}}{{\mathcal{F}}_{1}}-
\frac{\Omega_{2}\sin{2\theta}}{2{\mathcal{F}}_{3}}-
\frac{\Omega_{2}(\widetilde{\omega}_{y}+
\widetilde{\omega}_x)\sin^2{2\theta}\sinh{\Omega_{1}t}}
{4{\mathcal{F}}_{1}{\mathcal{F}}_{3}}
\nonumber \\
&
-\frac{\Omega_{1}(\widetilde{\omega}_{y}+
\widetilde{\omega}_x)^2\sin^2{2\theta}\cos^2{\theta}
\sinh{\Omega_{1}t}\sin{\Omega_{2}t}}
{4{\mathcal{F}}_{1}^2{\mathcal{F}}_{3}}\biggr],
\nonumber\\
C(t)=&-\frac{m_{y}}{2\hbar}\biggl[
\frac{{\mathcal{F}}_{2}}{{\mathcal{F}}_{1}}\sin^2{\theta}+
\frac{{\mathcal{F}}_{4}}{{\mathcal{F}}_{3}}\cos^2{\theta}+
\frac{\Omega_{1}\Omega_{2}}{2{\mathcal{F}}_{1}
{\mathcal{F}}_{3}}(\widetilde{\omega}_{y}+\widetilde{\omega}_{x})\sin^2{2\theta}
\nonumber\\
&+\frac{\Omega_{1}^2(\widetilde{\omega}_{y}+
\widetilde{\omega}_{x})^2\sin^2{2\theta}\sin^2{\theta}
\sin{\Omega_{2}t}}{4{\mathcal{F}}_{1}^2{\mathcal{F}}_{3}}\biggr],
\nonumber\\
D(x,t)=&\frac{x\sin{2\theta}}{\hbar}\sqrt{m_{\rm P}m_{y}}\biggl(\frac{\Omega_{1}\Omega_{2}}{2{\mathcal{F}}_{1}
{\mathcal{F}}_{3}}(\widetilde{\omega}_{y}+\widetilde{\omega}_{x})\cos{2\theta}+
\frac{{\mathcal{F}}_{2}}{{\mathcal{F}}_{1}}
+ \frac{{\mathcal{F}}_{4}}{{\mathcal{F}}_{3}}
\nonumber\\
&
+\frac{\Omega_{1}^2(\widetilde{\omega}_{y}+
\widetilde{\omega}_{x})^2\sin^2{2\theta}\sin{\Omega_{2}t}}{8{\mathcal{F}}_{1}^2{\mathcal{F}}_{3}}\biggr)
\nonumber\\
&
+\frac{i\alpha^{*}}{2}\sqrt{\frac{2m_{y}\widetilde{\omega}_{x}}{\hbar}}\sin{2\theta}
\biggl(-\frac{\Omega_{2}}{{\mathcal{F}}_{1}{\mathcal{F}}_{3}}(\widetilde{\omega}_{y}
+\widetilde{\omega}_{x})\cos^2{\theta}\sinh{\Omega_{1}t}-\frac{\Omega_{2}}{{\mathcal{F}}_{3}}
+\frac{\Omega_{1}}{{\mathcal{F}}_{1}}
\nonumber\\
&
-\frac{\Omega_{1}(\widetilde{\omega}_{y}+\widetilde{\omega}_{x})^2
\sin^2{2\theta}\sinh{\Omega_{1}t}\sin{\Omega_{2}t}}
{4{\mathcal{F}}_{1}^2{\mathcal{F}}_{3}}\biggr).\lb{ft2}
\end{align}

The full state \eqref{sol1} describes an entangled state between this
quantum black hole and its Hawking radiation toy model. One can then
calculate from it the
reduced density matrix for the black hole by integrating out
the $y$-part.
Tracing out the Hawking radiation thus
gives a mixed state for the analogue of the black hole itself \cite{DK}.
In fact, the presence of entanglement between
the black hole state and the Hawking radiation
is of particular interest as emphasized, for example, in
\cite{Zeh05}. It is important to recall that the black hole is an open
system, which by itself (without taking into account the Hawking
radiation as well as all other fields interacting with it) does not evolve
unitarily. The information-loss problem for black holes can only refer
to the closed system of black hole plus all other degrees of freedom
which are entangled with it. By our very ansatz, in our model
the full evolution of
black hole plus Hawking radiation is unitary with respect to
semiclassical time.

The diagonal element of the reduced density matrix for
the black hole is computed from
\be
\lb{rm1}
\rho_{xx}={\rm tr}_{y}\rho=\int|\langle x,y|x,y\rangle|^2 dy,
\ee
where we use $|x,y\rangle \equiv \psi(x,y,t)$.
Inserting \eqref{sol1} into \eqref{rm1}, one finds,
\be
\lb{rm2}
\rho_{xx}=|F|^2\exp{\biggl( x^2(A+A^{*})+x
  (B+B^{*})\biggr)}\sqrt{-\frac{\pi}{C+C^{*}}}
\exp{\biggl(\frac{(D+D^{*})^2}{4(C+C^{*})}\biggr).}
\ee
Similarly, the off-diagonal element of the reduced density matrix for
the black hole becomes
\be
\lb{rm3}
\rho_{xx'}=|F|^2\exp{\biggl( x^2 A+ x'^2 A^{*}+x B+x'
  B^{*}\biggr)}\sqrt{-\frac{\pi}{C+C^{*}}}
\exp{\biggl(\frac{(D(x)+D^{*}(x'))^2}{4(C+C^{*})}\biggr)},
\ee
where the sum $D(x)+D^{*}(x')$ implements an additional dependence on $x$
and $x'$ as we can verify from (\ref{ft2}). Here we shall not discuss
further the non-diagonal elements, which describe decoherence effects \cite{DK}.

Figure 4 depicts $\rho_{xx}$ for $\mu=0$, thus representing a
situation where there is no explicit back-reaction term.
The parameters are specified in order to have $m_{\rm P}\gg m_y$;
this seems to be a reasonable choice, since the energy scale of
Hawking radiation is much smaller than the Planck mass, except perhaps
in the
very last stage of the evaporation.
We can observe that Figure 4 is essentially identical with Figure 2
(obtained without taking into acount $f(x_0,p_0,t)$).
In fact, having $\mu = 0$ corresponds to take $\theta = 0$ in
\eqref{ft2} and \eqref{rm2};
consequently, we obtain the solution \eqref{coh1}, since
$\lim\limits_{\theta\rightarrow0}\rho_{xx}$ =
$\vert\psi_x^{\alpha}(x,t)\vert^2$.

\begin{figure}[h]
\setlength{\belowcaptionskip}{15pt}
\renewcommand{\baselinestretch}{1.2}
  \centering \includegraphics[width=5.5in]{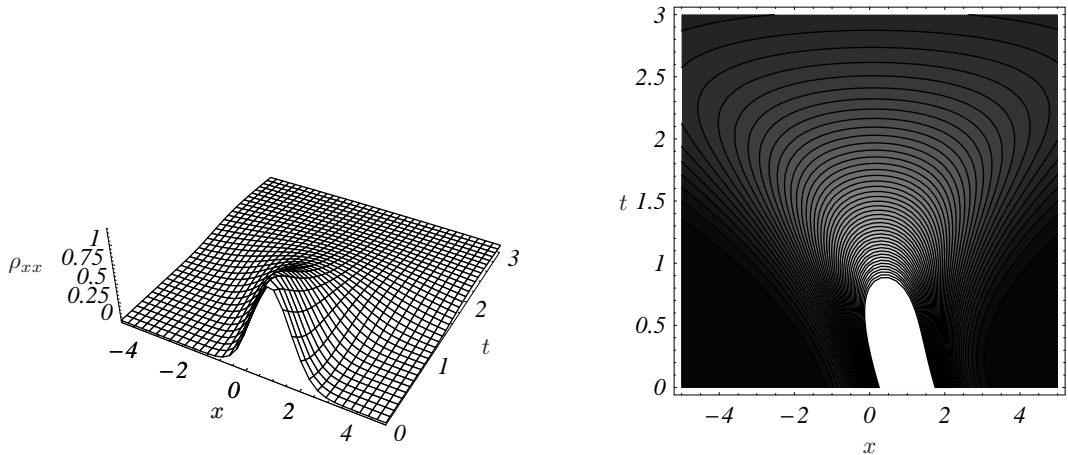}
 \caption{{\protect \footnotesize Evolution of $\rho_{xx}$, with
     $m_{\rm P}=\hbar=\omega_x=x_{0}=1$ and  $p_{0}=-1$; $t_0=0$ for simplicity,
     and $\mu=0$ (no back-reaction coupling), $\omega_y=\omega_x\times
     10^{5/2}$, $m_y=m_{\rm P} \times 10^{-5}$. In the contour plot the
     brighter areas correspond to higher values for $\rho_{xx}$.}}
\end{figure}

The most interesting situation, however, is to allow $\mu\neq0$
in order to have an idea of the effect of back reaction on $\rho_{xx}$.
Figure 5 shows what happens for $\mu$ varying from 0
to 100.
In fact, significant modifications to $\rho_{xx}$ emerge mostly when $\mu$
becomes larger than 1:
we can observe a strong modification of $\rho_{xx}$ when
the back reaction term comes into play.
Due to the entanglement between the black hole state and the Hawking radiation,
$\rho_{xx}$ is now sensitive
to modifications of $\omega_y$ or $m_{y}$.
The particular choices of these parameters,
which constitute the information carried by Hawking radiation, lead to
very different forms of $\rho_{xx}$.

One recognizes from Figure~5 that there is for large $\mu$, that is,
for large back reaction, a supression of the strong squeezing. There
remains a relatively narrow wave packet whose width, however, strongly
oscillates. 

\begin{figure}[h]
\setlength{\belowcaptionskip}{15pt}
\renewcommand{\baselinestretch}{1.2}
  \centering \includegraphics[width=5.8in]{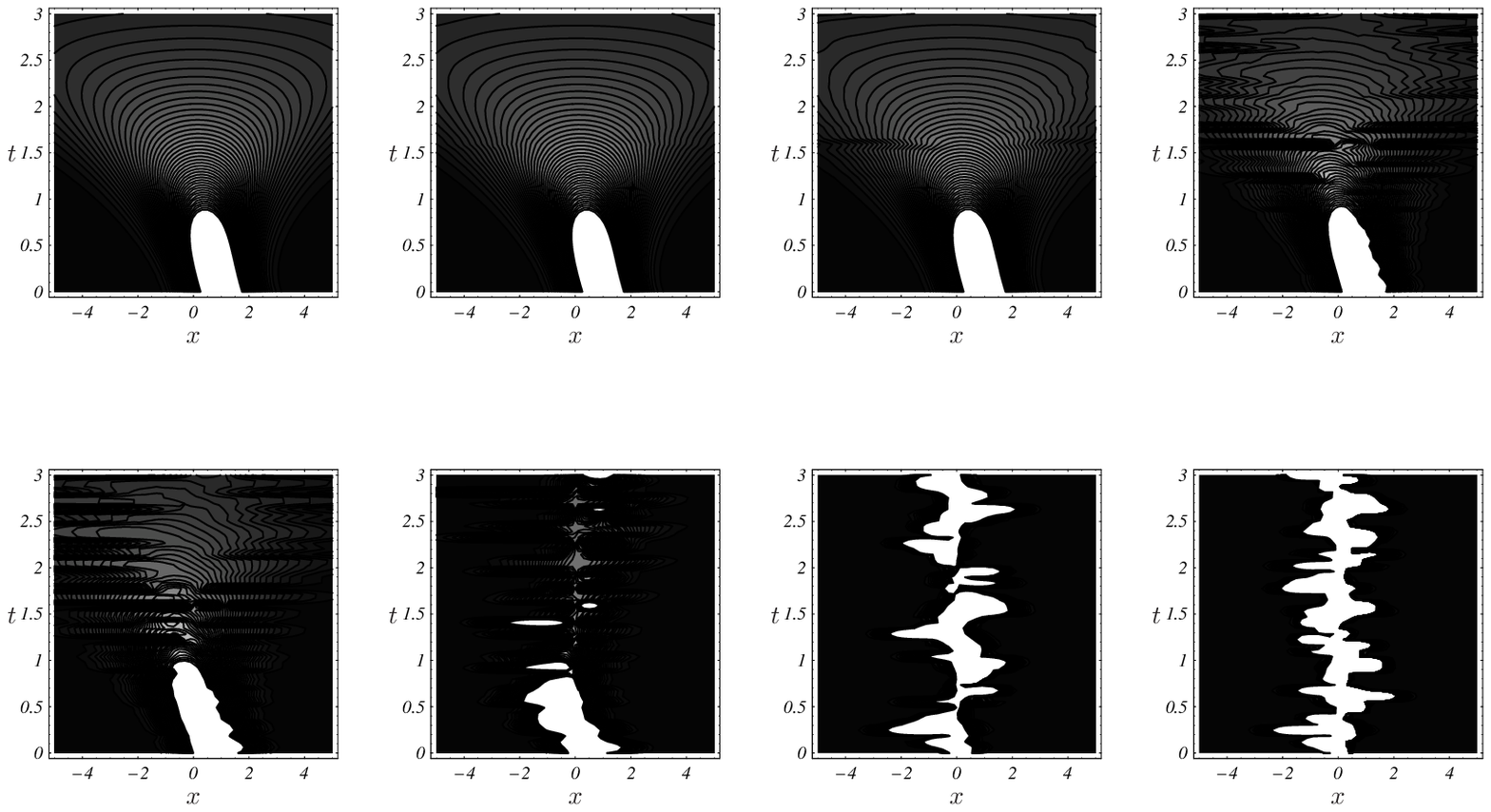}
 \caption{{\protect \footnotesize Time evolution of $\rho_{xx}$, with
     $m_{\rm P}=\hbar=\omega_x=x_{0}=1$; $t_0=0$ and $p_{0}=-1$ for simplicity,
     and $\mu$ (graphics from left to right and top to bottom)
     assuming the values of the set $\{0, 0.5, 1, 5, 10,
     20, 50, 100\}$ , $\omega_y=\omega_x\times 10^{5/2}$, $m_y=m_{\rm P} \times
     10^{-5}$. In the contour plot the brighter areas correspond to
     higher values for $\rho_{xx}$.}} 
\end{figure}

Instead of tracing out Hawking radiation from the full entangled state
\eqref{sol1}, we can trace out the black-hole state and thereby
arrive at the density matrix $\rho_{yy}$. The
computation of $\rho_{yy}$
processes in a similar fashion as for $\rho_{xx}$, and the formal
expression for it
is comparable to \eqref{rm2}. The interest in this quantity arises from
the fact that
the effect of the back reaction between the black hole and the Hawking
radiation
(in the context of this simplified model) can be explored from what
effectively leaves
the hole. In particular, the issue of information being carried by
this radiation
is of special interest. The question we may ask is what should
we expect
to detect if the Hawking radiation were taken to be a signal affected
by back reaction
effects. In Figure 6 we represent $\rho_{yy}$ for various values of
$\mu$. We recognize that the diagrammes for $\rho_{xx}$ and
$\rho_{yy}$ look similar for large $\mu$. This could be due to the
large entanglement between the black hole and its Hawking radiation,
which leads to similar reduced density matrices. 
\begin{figure}[h]
\setlength{\belowcaptionskip}{15pt}
\renewcommand{\baselinestretch}{1.2}
  \centering \includegraphics[width=5.8in]{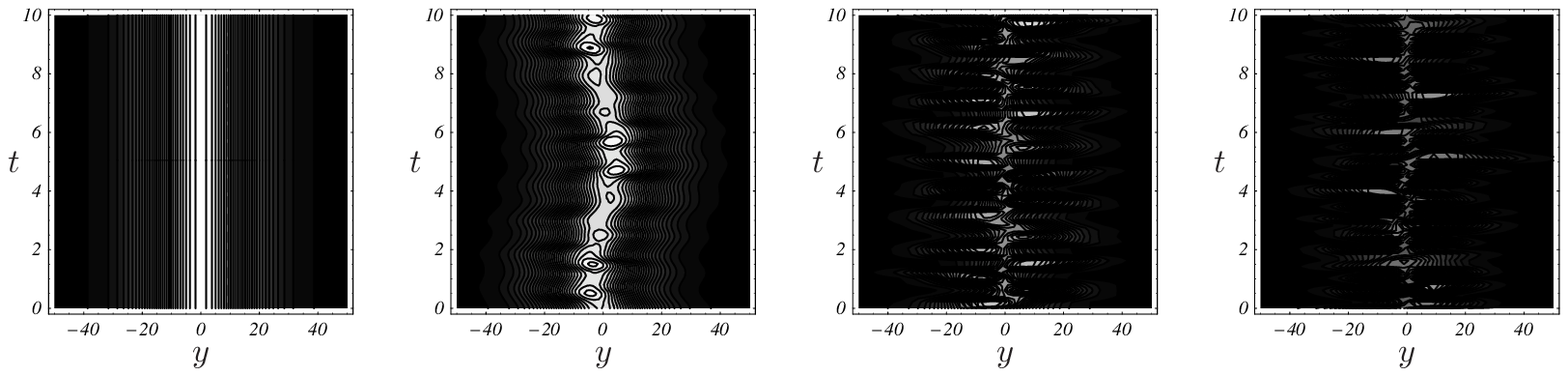}
 \caption{{\protect \footnotesize Time evolution of $\rho_{yy}$, with
     $m_{\rm P}=\hbar=\omega_x=x_{0}=1$ and $p_{0}=-1$; $t_0=0$ for simplicity,
     and $\mu$ (graphics from left to right)
     assuming the values of the set $\{0, 1, 5, 10\}$,
     $\omega_y=\omega_x\times 10^{5/2}$, $m_y=m_{\rm P} \times
     10^{-5}$. In the contour plot the brighter areas correspond to
     higher values for $\rho_{yy}$. For large values of $\mu$ the
     results look qualitatively similar to the results for $\rho_{xx}$
     depicted in Figure~5. If the back reaction is large, the
     difference between the black hole and the Hawking radiation
     begins to disappear.}}
\end{figure}

\section{Summary and conclusion}

The main purpose of this paper is to get an intuitive, though
tentative, insight into how
the final phase of black-hole evaporation may look like. In the
geometrodynamical framework used here, the question is:
suppose we have solved the quantum constraints and found the
wave function describing the black hole and its radiation -- which
characteristics would identify it if some quantum state were detected?

We have argued that in addressing this question we have to deal not
with the full quantum constraints of the Universe, but with a
(functional) Schr\"odinger equation which contains the exact
Hamiltonian of the quantum black hole together with the semiclassical
time $t$ of the rest of the Universe. After all, there are observers
who observe the quantum black hole from outside and who have a clock
at their disposal.

We have constructed and investigated
a very simple model: two coupled harmonic oscillators which mimic
black hole and Hawking radiation. We have found that our analogue of
a black-hole
state experiences a strong squeezing during evaporation, but that this
squeezing may disappear for large back reaction; in this limit the
reduced density matrices for black hole and Hawking radiation look
similar to each other -- their differences begin to disappear. 

The simplicity of our model allows at best to give some heuristic
insight. It is most unlikely that the end phase of a black hole can
be described by a system as simple as coupled oscillators. However,
it is possible that some of the aspects discussed here -- high squeezing
of the black-hole state and large entanglement with Hawking radiation --
occur also for the realistic quantum states. Moreover, features of an
oscillator system are often discussed even in full approaches to
quantum gravity, such as the recent proposal for a flux-area operator
with an equidistant eigenvalue spectrum \cite{Barbero}. 

A big open question is the relation of such simplified models like
ours to full field-theoretic approaches. This question can, of course,
only be answered from the full theory, not from the angle of our
model. But there exists an abundance of papers in the literature which
address quantum aspects of black holes from this `minisuperspace'
point of view, with \cite{kuchar,LWH,KL,BK,others} listing only some
of them. The hope there is always that one or the other result will
survive in the full elusive description. Our paper is written in the
same spirit as these papers.

Many more questions remain to be answered. One concerns the calculation of
the black-hole entropy and the recovery of the Bekenstein--Hawking
formula in the limit of large black holes. Great progress has been
achieved here in loop quantum gravity \cite{OUP,Rovelli} and string
theory \cite{Zwiebach}, but some results have also been obtained from
the Wheeler--DeWitt equation in quantum geometrodynamics, cf.
\cite{Vaz2} and the references therein. It would be of interest to
give some exact results in the context of our simplified model. One
must not forget that a black hole is a genuine open quantum system
because it is susceptible to even small interactions with other fields
(including its own Hawking radiation, as discussed here)
\cite{Zeh05}. The process of decoherence would thus play a crucial
role in the discussion \cite{deco}. This is especially important
because it is known from quantum mechanics that squeezed states -- and
the final black-hole state in our model approaches such a state
for not too large back reaction -- are
highly sensitive to decoherence. A possible connection with
observation could be made when addressing primordial black holes --
small relics from the early Universe, which in the appropriate mass
range could be evaporating in the present phase of the Universe
\cite{Carr}.
We hope to address some of these issues in future publications.


\section*{Acknowledgements}
This work was supported by DAAD-GRICES/2004/2005 - D/03/40416,
CRUP-AI-A-21/2004, POCI/FIS/57547/2005.
 C.K. acknowledges kind hospitality at the Universidade
da Beira Interior, Covilh\~a, Portugal, where this work was begun.



\end{document}